\documentclass[reprint,amsmath,amssymb,aps]{revtex4-2}

\usepackage{graphicx}
\usepackage{dcolumn}
\usepackage{bm}
\usepackage{gensymb}
\usepackage[colorlinks=true,linkcolor=blue, citecolor=blue]{hyperref}
\usepackage{xcolor}
\graphicspath{{./images/}}

\begin{document}
	
	
	\title{Photoinduced phase transitions and lattice deformation in 2D NbOX\ensuremath{_{2}} (X=Cl, Br, I)}

\author{Carmel Dansou\textsuperscript{1}, Charles Paillard\textsuperscript{1,2}, Laurent Bellaiche\textsuperscript{1,3}}
\affiliation{\textsuperscript{1} Smart Ferroic Materials center, Institute for Nanosciences \& Engineering and Department of Physics, University of Arkansas, Fayetteville AR 72701, USA.}
\affiliation{\textsuperscript{2}Université Paris-Saclay, CentraleSup\'{e}lec, CNRS, Laboratoire SPMS, 91190 Gif-sur-Yvette, France.}
\affiliation{\textsuperscript{3}Department of Materials Science and Engineering, Tel Aviv University, Ramat Aviv, Tel Aviv 6997801, Israel.}

	\date{\today}
	
	\begin{abstract}
		We present a comprehensive investigation of light-induced phase transitions and strain in two-dimensional NbOX\ensuremath{_{2}} (X = Cl, Br, I) using first-principles calculations. In particular, we identify a light-induced ferroelectric-to-paraelectric phase transition in these 2D systems. Furthermore, we demonstrate the possibility of inducing an antiferroelectric-to-paraelectric transition under illumination. Additionally, we find that these 2D systems exhibit significant photostrictive behavior, adding a new functionality to their already notable optical properties. The ability to control and manipulate ferroelectric order in these nanoscale materials through external stimuli, such as light, holds considerable promise for the development of next-generation electronic and optoelectronic devices.
	\end{abstract}
	
	\maketitle
\section{Introduction}
The discovery of materials capable of changing shape in response to external stimuli, such as piezoelectrics and piezomagnets, has led to the development of actuators, micro-electromechanical systems, and microrobotics \cite{sekhar2021piezoelectricity,shaukat2023piezoelectric,berlincourt1964piezoelectric,mohith2020recent}. Like piezoelectricity and piezomagnetism, photostriction \cite{ku20,finkel}--a phenomenon in which light induces mechanical strain in a material--enables to use light as an external handle for next-generation wireless and remote-controlled devices \cite{poos,sun}. Consequently, in the last decade, intensive research has been conducted to understand photostriction and new functional materials such as ferroelectrics \cite{Haleoot2017,Paillard2017,Paillard2016,matz,gao2023}, multiferroics \cite{kund,kundlig,wen2013,schick2014,pail1,gu2023}, organic polymers \cite{naka}, inorganic semiconductors \cite{busch} and halide perovskites \cite{zhou20, paillard2023,bpeng20}.

The niobium oxide dihalides family,  NbOX\ensuremath{_{2}} (X = Cl, Br, I), has recently gained a lot of attention as a new class of layered FE materials \cite{jia_niobium_2019,rijnsdorp_crystal_1978,liu_ferroelectricity_2023,wu_data-driven_2022} with potential applications as quantum light sources \cite{guo_ultrathin_2023} and sensors thanks to their tunable bandgap, monolayer-like excitonic behavior even in bulk form \cite{wang_indirect_2024}, and efficient second-harmonic generation (SHG) response~\cite{abdelwahab_giant_2022}. Moreover, NbOCl\ensuremath{_{2}} has been reported to possess high carrier mobility \cite{zh_2024} and strong light absorption in the visible range \cite{zhang_linear_2024}, making it a promising material for optoelectronic applications. Recent studies have also shown that epitaxial strain can significantly enhance the electro-optic behavior of both 2D and 3D NbOI\ensuremath{_{2}} \cite{zhang_giant_2024}.

Motivated by these observations of light-matter coupling in two-dimensional NbOX\ensuremath{_{2}} (X = Cl, Br, I), we employed Density Functional Theory (DFT) calculations to investigate the possibility of inducing phase transition using optical excitation as well as light-induced strains in the NbOX\ensuremath{_{2}} family. Through systematic simulation of optical excitation in the paraelectric (PE), ferroelectric (FE), and antiferroelectric (AFE) phases of 2D-dimensional NbOX\ensuremath{_{2}}, we show that a FE-PE transition as well as an AFE-PE transition can be optically induced. Additionally, we predict that 2D NbOX\ensuremath{_{2}} structures exhibit remarkable photostrictive properties, therefore showing their potential application in light-controlled devices.

\section{Methods and ground state structures}

In this study, DFT calculations were performed using the open-source code Abinit \cite{ab}. Optimized Norm-Conserving Vanderbilt pseudopotentials \cite{haman} are employed in our simulations. To approximate the exchange-correlation energy, we used the generalized gradient approximation proposed by Perdew, Burke, and Ernzerhof (PBE) \cite{pbe}.  A plane wave cut-off energy of 40 Hartree was used for the three structures, and integration over the Brillouin zone was performed using a non-shifted $6\times 12\times 1$ k-mesh. The density during self-consistent field (SCF) calculations was considered converged when the residual forces on the structure were smaller than $10^{-6} \text{ Ha.Bohr}^{-1}$. Single layer NbOX\ensuremath{_{2}} was simulated in the present study. To eliminate spurious interactions between periodic images in the $z$-direction, a vacuum layer of 20 \AA~ was introduced between periodic image.

To simulate above-bandgap optical excitation, we used DFT with the constrained occupation scheme, recently developed to study light-induced phenomena in ferroelectric and other materials \cite{paillard2019}. In this constrained DFT (cDFT) approach, photoexcitation is modeled by constraining $n_{ph}$ electrons in the conduction band and $n_{ph}$ holes in the valence band during each SCF cycle. The excited electrons and holes are treated as two Fermi-Dirac distributions with their own quasi-Fermi level. A smearing of 0.004~Ha was applied. The system is relaxed under this constraint until convergence is achieved, yielding an SCF solution for the two chemical potentials.

After full relaxation in dark, the FE and the PE phases adopt orthorhombic structures with $Pmm2$ (No. 25) and $Pmmm$ (No. 47) space group symmetry, respectively, while the AFE phase adopts a monoclinic structure with $P2/m$ (No. 10) space group symmetry. These structures are consistent with previous DFT and experimental studies \cite{wu_data-driven_2022,jia_niobium_2019,wang_anisotropic_2024,zhang_structural_2024,mohebpour_origin_2024}. Vesta illustrations of the final relaxed structures are shown in the Supplemental Materials (SM) \cite{xx}. In the FE phase, polarization is induced by the off-center displacement ($d$) of the Nb atoms along the $b$-axis. The PE and AFE phases are not polar and have zero total polarization. However, in the AFE phase, the Nb atoms in adjacent cells displace in opposite direction along the $b$-axis. This gives rise to a null {\it overall} polarization but non-zero {\it local} polarizations in the AFE phase. As we will show later, due to the presence of local polar order in the AFE state, its interaction with optical excitation significantly differs from that of the nonpolar PE phase.

\section{Photo-induced phase transition in 2D NbOX\ensuremath{_{2}}}
\begin{figure}[ht!]
	\centering
	\includegraphics[width=1\linewidth]{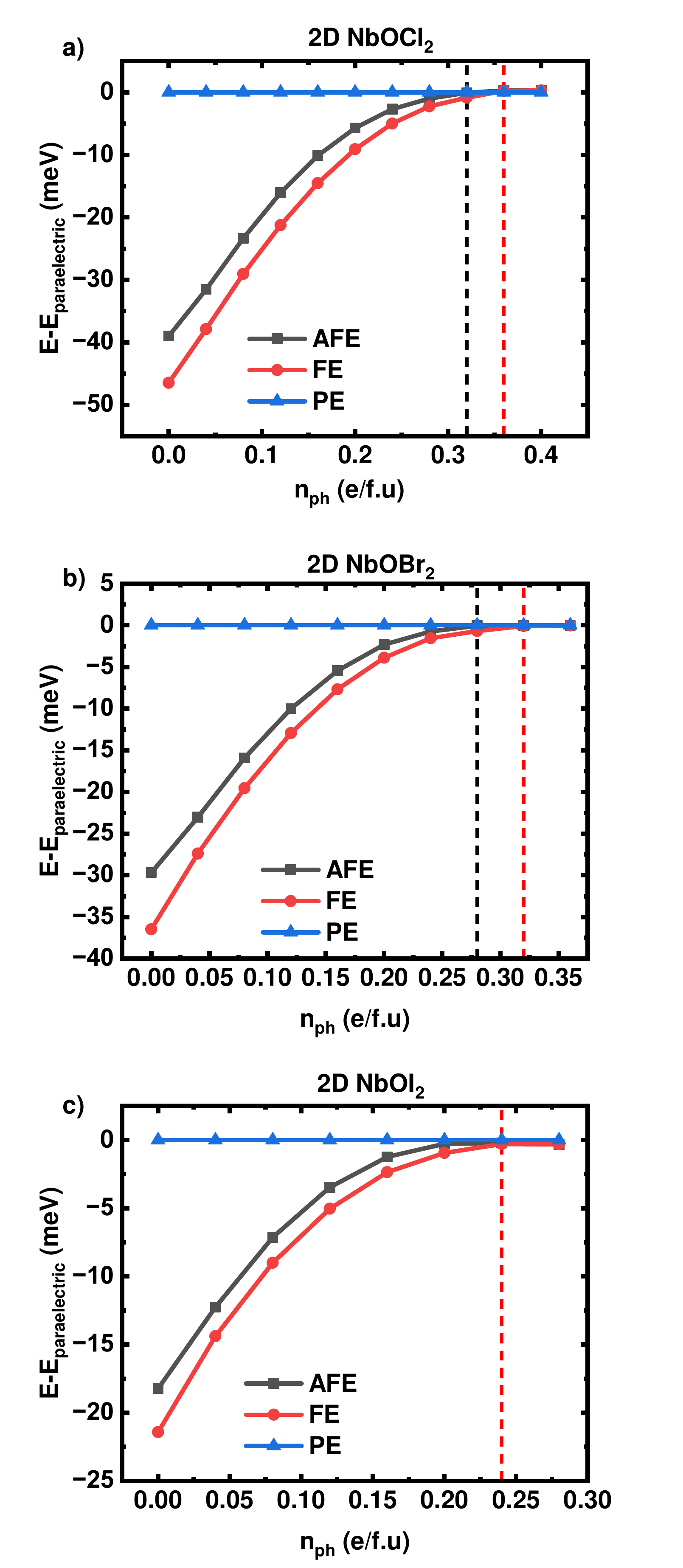}
	\caption{\label{f1e} Energy of the PE, FE and AFE phases of each of the 2D NbOX\ensuremath{_{2}} materials. a) 2D NbOCl\ensuremath{_{2}}, b) 2D NbOBr\ensuremath{_{2}} and c) 2D NbOI\ensuremath{_{2}}. The energy of the PE phase is set to zero. The black and red dashed lines in a) and b) are guide for the eye to show the transition from AFE to PE and the transition from FE to PE.}
\end{figure}

We start by looking at the photo-induced phase transition in the considered 2D NbOX\ensuremath{_{2}} materials. The total energy of the three phases for  NbOCl\ensuremath{_{2}}, NbOBr\ensuremath{_{2}} and NbOI\ensuremath{_{2}}, as a function of the number of photoexcited carriers concentration is shown on Fig \ref{f1e}. We set the PE phase as the origin of energy at each value of $n_{ph}$. As stated above, in the dark, the FE is the ground state for all the  NbOX\ensuremath{_{2}} materials, followed by the AFE phase and the PE at a much higher energy. As noted in previous studies \cite{jia_niobium_2019,wu_data-driven_2022}, although the AFE phase is not the ground state, it is a metastable phase very close in energy to the FE phase. As we increase $n_{ph}$, we notice a gradual reduction in the energy difference between the FE and PE phases but also between the AFE and PE states. As can be seen from Fig \ref{f1e}a, for the 2D NbOCl\ensuremath{_{2}}, as we increase $n_{ph}$, the AFE phase transforms into the PE phase at $n_{ph}=0.32$ e/f.u. Subsequently at $n_{ph} = 0.36$ e/f.u, the FE phase transforms into the PE. Beyond $n_{ph}=0.36$ e/f.u, only the PE phase exists. The  $n_{ph}$ for the AFE-PE and FE-PE  transitions correspond to $2.14\times 10^{14}~ \text{cm}^{-2}$ and $4.8\times 10^{14}~ \text{cm}^{-2}$ excited electron-hole pairs with a typical $2$ eV pump at $I_{0} = 5.0~ \text{mJ.cm}^{-2}$ and $I_{0} = 6.0~ \text{mJ.cm}^{-2}$ fluences, respectively (the details of this estimation is shown in the SM \cite{xx}). For the 2D NbOBr\ensuremath{_{2}}, the AFE and the FE phases become paraelectric at $n_{ph} = 0.28$ e/f.u and $n_{ph} = 0.32$ e/f.u, respectively. These are equivalent to exciting about $3.4\times 10^{14}~ \text{cm}^{-2}$ and $3.8\times 10^{14}~ \text{cm}^{-2}$ electron-hole pairs with the same $2$ eV pump at $I_{0} = 4.0~ \text{mJ.cm}^{-2}$ and $I_{0} = 4.50~ \text{mJ.cm}^{-2}$ fluences, respectively. And finally for the 2D NbOI\ensuremath{_{2}}, the FE and the AFE states are all driven to the PE phase at approximately $n_{ph} = 0.24~$ e/f.u corresponding to a density of excited electron-hole pairs of about $1.3\times 10^{14}~ \text{cm}^{-2}$ with a typical $2.0$ eV pump with $I_{0} = 3.0~ \text{mJ.cm}^{-2}$ as fluence. Beyond $n_{ph}=0.24$ e/f.u, all the three phases of NbOI\ensuremath{_{2}} evolve as a single PE phase. Using the online FINDSYM utility \cite{findsym}, we also follow the symmetry of each phase as a function of $n_{ph}$ e/f.u. Consistently with the transition observed in the energy diagram, for each of the three considered 2D allotropes, the FE and the AFE states transform from the $Pmm2$ and $P2/m$, respectively to the centrosymmetric  $Pmmm$ space group of the PE phase. We did not, however, observe any transition between the FE and the AFE phases.
The transition from FE to PE phases in  FE materials has been a subject of previous discussions in the literature. It was found that photoexcited charges can alter polar order in classical perovskite FE materials such as BaTiO\ensuremath{_{3}}, PbTiO\ensuremath{_{3}}, causing a structural reorganization and  suppressing ferroelectric instability, thereby causing transitions from a FE to a PE phase \cite{paillard2019,krapivin202}. We follow the evolution of polarization in the FE phases as function of $n_{ph}$. The uniaxial in-plane polarization in each of the NbOX\ensuremath{_{2}} materials is estimated using the Born effective charges formula $\mathbf{P}_{y} = S^{-1} \sum_{i} Z_{yy}^{i}\mathbf{u}_{y}$ with $Z_{yy}^{i}$  is the $y$ component of the Born effective charge tensor  (See SM \cite{xx} for their values obtained using density functional perturbation theory (DFPT) calculations) of the $i$ atoms, $\mathbf{u}_{i}$ its displacement with respect to the centrosymmetric position and $S$ the in-plane area of the simulated cell.
\begin{figure}[ht!]
	\centering
	\includegraphics[width=1\linewidth]{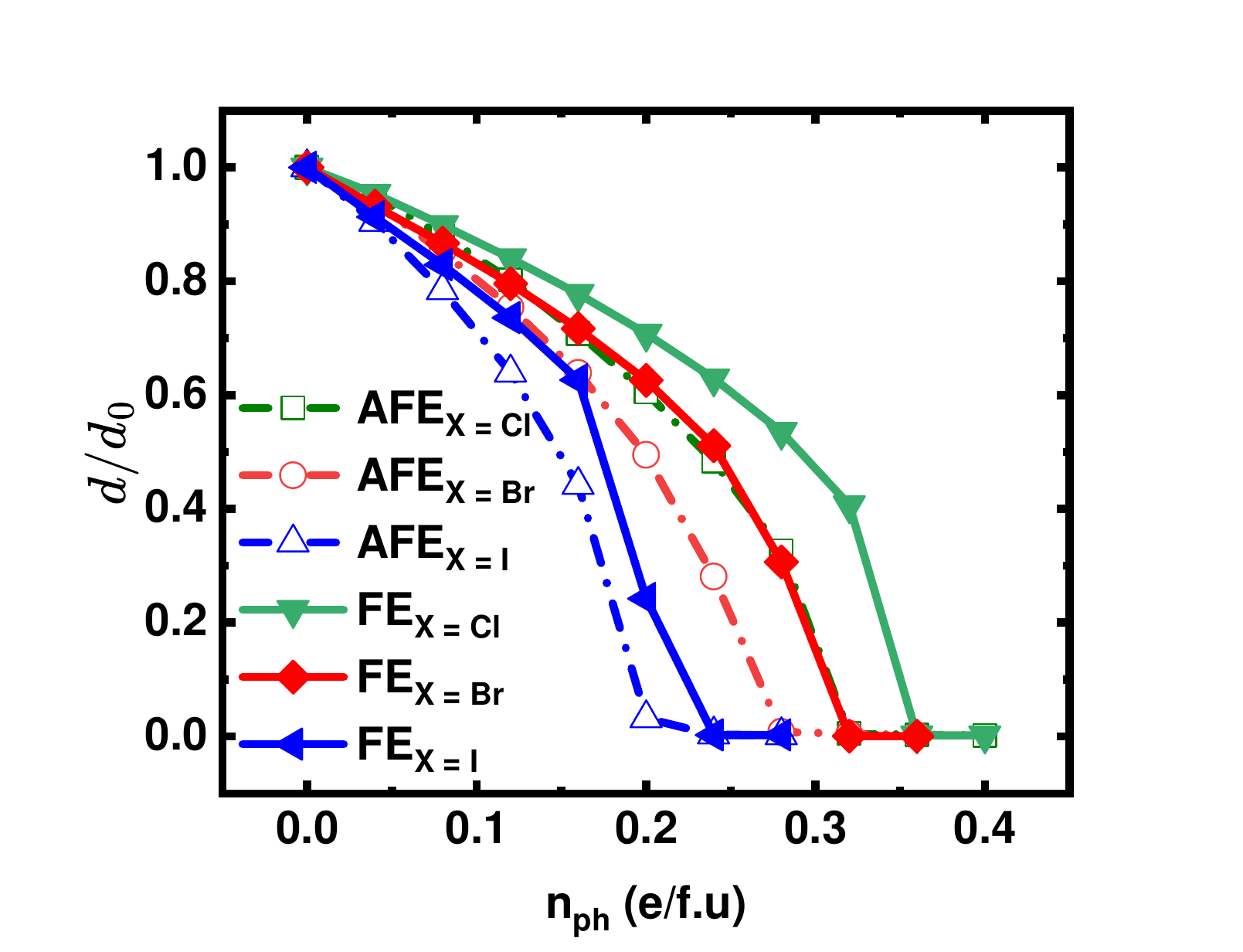}
	\caption{\label{f2e} a) Evolution of the ratio $d/d_{0}$ of the FE and AFE phases of each of the 2D NbOX\ensuremath{_{2}} materials as function of $n_{ph}$: green line for 2D NbOCl\ensuremath{_{2}}, red line for 2D NbOBr\ensuremath{_{2}} and blue line for 2D NbOI\ensuremath{_{2}}. Dashed linse and open symbols are for the AFE phase. Full lines and filled symbols for the FE phase.}
\end{figure}
 We also map out the off-center displacement $d$ of two Nb atoms in our cell as function of $n_{ph}$. This parameter $d$ is known to drive the uniaxial polar behavior in NbOX\ensuremath{_{2}} \cite{jia_niobium_2019}. The ratio $d/d_{0}$ as function of $n_{ph}$ is shown on Fig \ref{f2e} as full line and filled symbols for the FE phase of each considered  2D NbOX\ensuremath{_{2}}, where $d_{0}$ is the displacement in dark ($n_{ph}=0$). The parameter $d$ is decreased by photoexcitation and eventually vanishes at the value of $n_{ph}$ where the FE phase transforms into the centrosymmetric PE phase for each 2D NbOX\ensuremath{_{2}} materials (the polarization as function of $n_{ph}$ is shown the SM \cite{xx}, follows the same trend and vanishes at the transition point). Previous studies on 2D FE materials such as SnS, SnSe, and similar materials have shown similar polarization behavior \cite{Haleoot2017}, but the transition to the PE phase has not been explicitly shown. The effect of illumination on the polarization in the FE phase lets us infer that the observed FE-PE transition is of the same nature as those previously discussed in bulk FE materials \cite{paillard2019,krapivin202}. 

On the other hand, the AFE phase lacks a global polar order. As such the total polarization of the system is always zero. To understand the origin of the light-induced transition from AFE to PE, we also focus on the evolution of the off-center displacement $d$ of Nb atoms in each of the two anti-polar cells of the crystal AFE structure. The ratios $d/d_{0}$, as function of $n_{ph}$ are shown in Fig \ref{f2e} as open symbols and dashed lines. The trend of $d/d_{0}$ indicates that free carriers decrease the off-center displacement of Nb atoms, which, like the local polarization (we estimate the local polarization by considering atoms in one of the polar sublattices of the unit cell \cite{xx}), vanishes at the value of $n_{ph}$ where the AFE phase transforms into the PE. This leads to the conclusion that, in the centrosymmetric AFE phase, light also screens the existing ``local dipoles" arrangement and thus leads to the transition observed in the energy diagram. This also means that we do not observe any FE-to-AFE or AFE-to-FE transition because the effect of illumination on the FE and AFE phases is intrinsically the same, that is screening of local electrical dipoles.

\section{Photostriction in the NbOX\ensuremath{_{2}}}
In this section, we discuss the photostrictive behavior of the 2D layered NbOX\ensuremath{_{2}} phases. In our simulations, we observe that the light-induced strains in the FE, AFE, and PE phases of the considered NbOX\ensuremath{_{2}} are qualitatively similar across structures; i.e., the PE phases exhibit similar qualitative behavior for all  $X$, as do the AFE and FE phases. Consequently, throughout this report, we focus on the results for NbOI\ensuremath{_{2}}, referring the reader to  the Supplemental Material \cite{xx} for data on 2D NbOCl\ensuremath{_{2}} and 2D NbOBr\ensuremath{_{2}} structures. Additionally, we report only the data in the linear regime ($n_{ph}= 0-0.16$) e/f.u for NbOI\ensuremath{_{2}}. The remaining data can be found in the SM \cite{xx}.
\begin{figure}[ht!]
	\centering
	\includegraphics[width=1\linewidth]{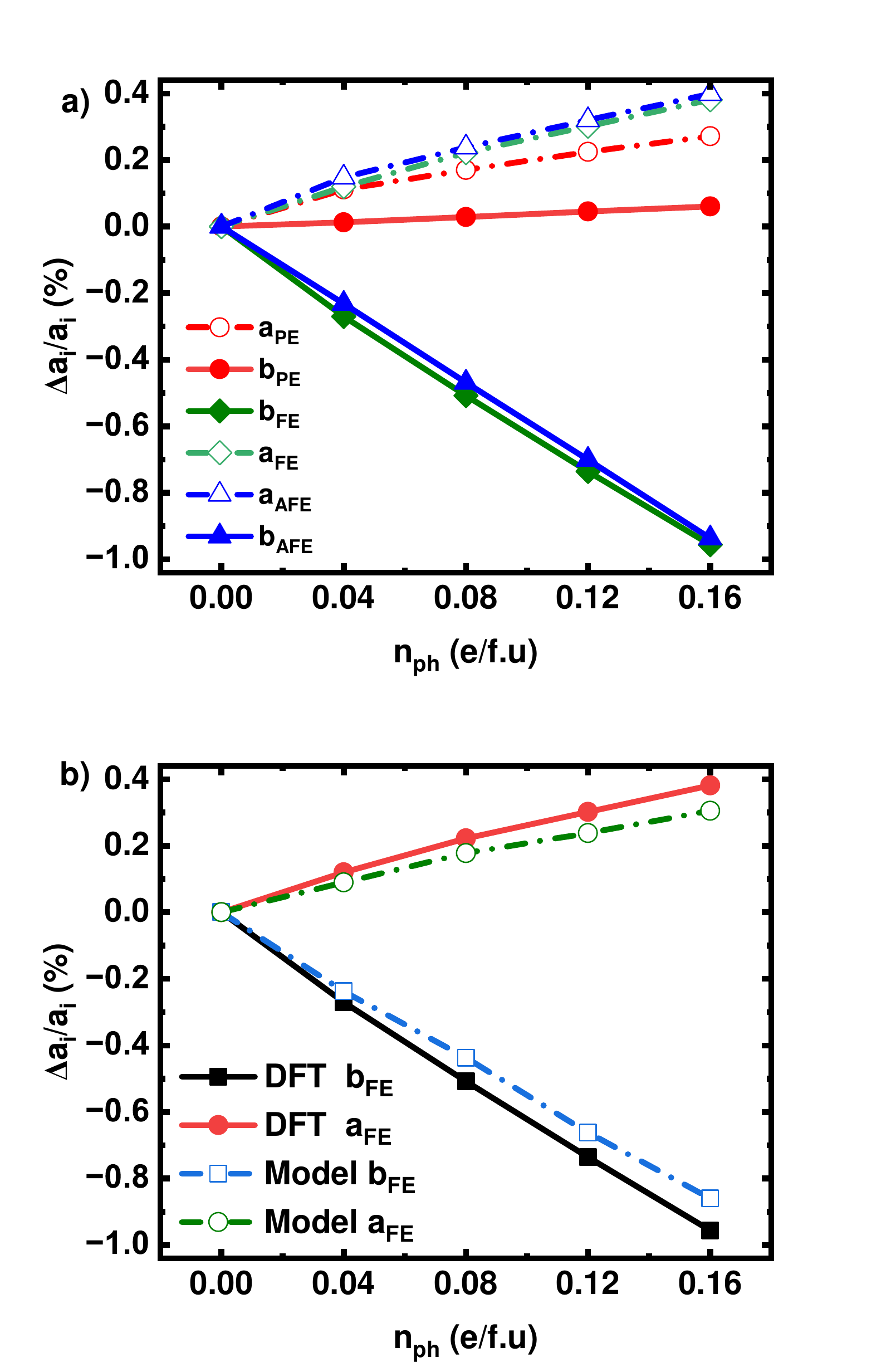}
	\caption{\label{f3e} a) Strain induced in the PE and AFE phases of the 2D NbOI\ensuremath{_{2}} materials as function of $n_{ph}=0-0.16$ e/f.u. Dashed line with open symbols are for strain in the nonpolar axis and full line are the polar/antipolar axis. Blue color for the AFE phase, red color for the PE phase and green color for the FE phase. b) DFT and converse piezoelectric model for the light-induced strains in the FE phases of the 2D NbOI\ensuremath{_{2}}}
\end{figure}

As shown on Fig \ref{f3e}a for 2D NbOI\ensuremath{_{2}}, in the FE and AFE phases of 2D NbOX\ensuremath{_{2}}, light induces an expansion along the nonpolar axis ($x$) and a compression along the polar/antipolar axis ($y$) similar to what was observed in early works on 2D FE (SnTe, SnS, GeS) \cite{Haleoot2017,luo_ultrafast_2023}. The same qualitative behavior was also predicted in bulk BiFeO\ensuremath{_{3}} (BFO), BaTiO\ensuremath{_{3}} (BTO)  and PbTiO\ensuremath{_{3}} (PTO) where it was found that illumination expands the nonpolar axis while compressing the polar axis \cite{pail1,pail2017}. This behavior is, however, in contrast with a recent study that showed that in weakly coupled SrTiO\ensuremath{_{3}}/PbTiO\ensuremath{_{3}} superlattices, the polar axis expands upon illumination \cite{datun}. As we explained above, the same qualitative behavior of AFE and FE phases can be understood in term of the screening of local polar order in the AFE state. Contrary to the observation in the AFE and FE phases, in the PE phase, both lattice lengths increase as we increase the number of photoexcited carrier in our simulations. This shows that the photo-induced strains is dependent on the phases of the layered 2D NbOX\ensuremath{_{2}}. 

To provide a measure of the photostrictive behavior in these 2D structures that could guide experimental exploration, we extract the slope strains vs photoexcited carrier concentration in the linear regime for each material's considered phases. This slope $\Delta\eta_{a}/ \Delta n_{ph}$ (\%/(e/f.u)) measures how much the lattice deforms per unit of photoexcited carrier in a formula unit of the material and allows for a better and guided comparison of the predicted photostriction with photostriction in other FE materials in the literature. The results are presented in Table \ref{tab0}.

\begin{table*}[ht!]
	\caption{Photostrictive coefficient $\Delta\eta_{a}/ \Delta n_{ph}$ (\%/(e/f.u)) of FE and AFE phases of 2D vdw NbOX\ensuremath{_{2}}}
	\centering
	\begin{tabular}{|c|  c c| c c  |c c |}
		\hline \hline
		\multicolumn{3}{|c| }{NbOCl\ensuremath{_{2}}}&\multicolumn{2}{ |c| }{NbOBr\ensuremath{_{2}}}& \multicolumn{2}{ |c |}{NbOI\ensuremath{_{2}}} \\ 
		\hline 
		& $\Delta\eta_{a}/ \Delta n_{ph}$  & $\Delta\eta_{b}/ \Delta n_{ph}$  &$\Delta\eta_{a}/ \Delta n_{ph}$  & $\Delta\eta_{b}/ \Delta n_{ph}$ &$\Delta\eta_{a}/ \Delta n_{ph}$  & $\eta_{b}/ \Delta n_{ph}$   \\\hline 
		AFE& 3.308& -4.903 & 3.325 &-5.168&2.428 & -5.856 \\
		
		FE & 3.272& -4.966 & 3.488&-5.314&2.363 & -5.944   \\ \hline \hline
	\end{tabular}
	\label{tab0}
\end{table*}

\begin{figure}[ht!]
	\centering
	\includegraphics[width=1\linewidth]{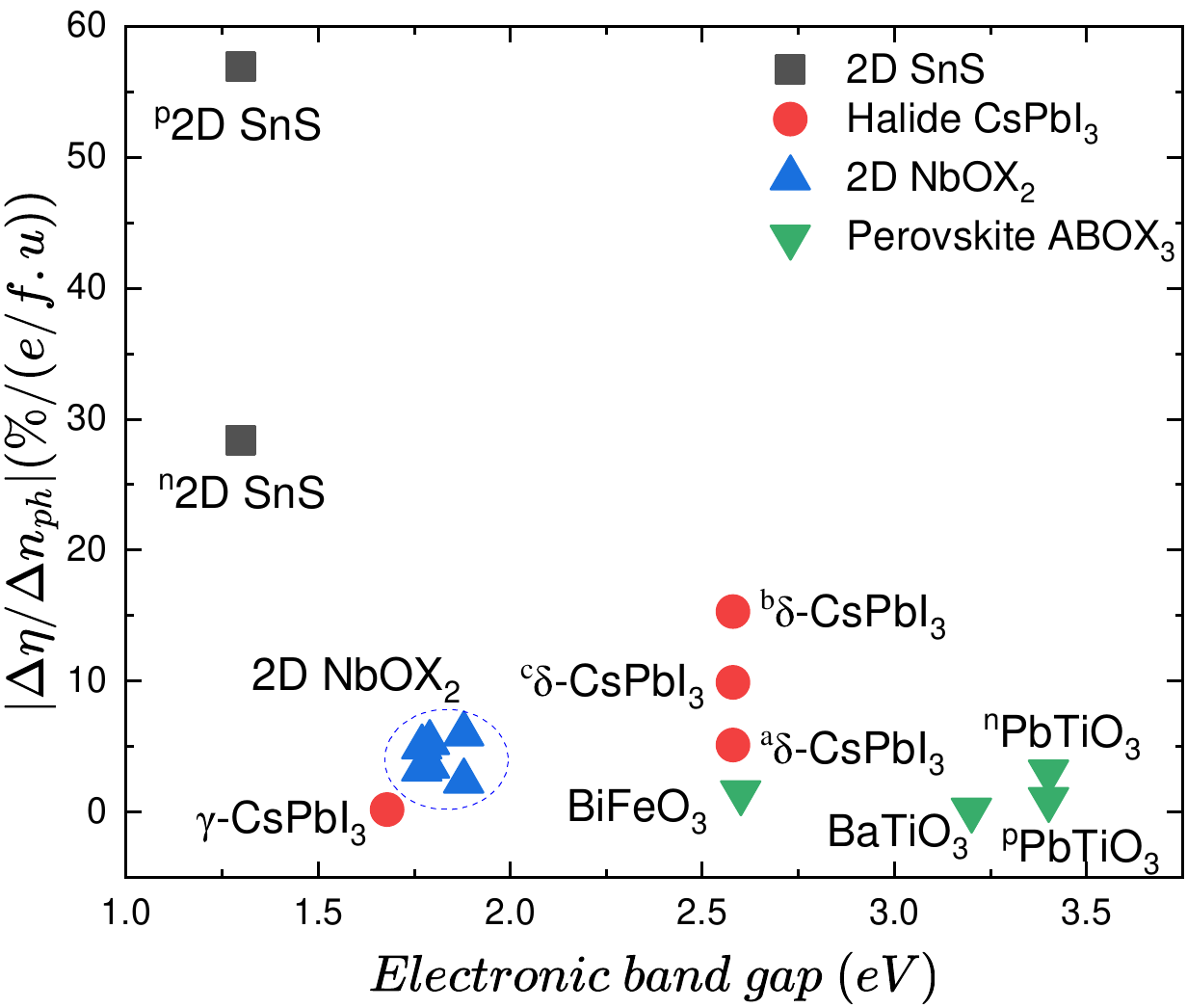}
	\caption{\label{f4e} Absolute values of predicted $\Delta\eta_{a}/ \Delta n_{ph}$ for various materials including classical oxide ferroelectrics, halide perovskites, other 2D materials and the prediction in 2D NbOX\ensuremath{_{2}} materials. The superscript before each element indicates the axis along which the deformation is measured: p represents the polar axis, n denotes the nonpolar axis, and a, b, and c correspond to the a-, b-, and c-axes, respectively}
\end{figure}

According to Table \ref{tab0}, in the linear regime, the FE and AFE phases exhibit nearly the same deformation along the polar/antipolar axis (the in-plane $b$-axis) as well as along the nonpolar axis (the in-plane $a$-axis). As compared to other classical bulk FE materials, the FE phase of NbOX\ensuremath{_{2}} show larger photostrictive behavior than \textit{R3c} BiFeO\ensuremath{_{3}} ($-1.5 $ \%/(e/f.u)), \textit{R3m} BaTiO\ensuremath{_{3}} ($-0.2 $ \%/(e/f.u)) and PbTiO\ensuremath{_{3}} ($-3.1 $ \%/(e/f.u)) \cite{paillard2023,pail1}. Along the nonpolar axis, they also deform more than PbTiO\ensuremath{_{3}} ($-1.0 $ \%/(e/f.u)) \cite{paillard2023,pail1}. They, however,  deform significantly less than 2D SnS ($-56.9 $ \%/(e/f.u)) \cite{Haleoot2017} along the polar axis. This comparative behavior is further illustrated in  Fig.\ref{f4e}.

Next, we look at the origin of the photostrictive behavior in the FE phase of these 2D materials. From previous experimental and theoretical studies \cite{Haleoot2017,pail1,Paillard2017,ku20,kundlig}, it has become a general consensus that the inverse piezoelectric effect is a large contribution to the photostriction phenomenon of ferroelectric materials. The excitation of thermalized photoexcited carriers alters polarization in the FE material, which can be thought of as if a light-induced internal electric field was generated in the material, subsequently leading to mechanical deformation through the inverse piezoelectric effect. 

In the following, we quantify the converse piezoelectric contribution to the observed photoinduced strains in the FE phases. Using the converse piezoelectric formula and following Refs. \cite{Paillard2017,Haleoot2017}, the light-induced strains can be related to the change in polarization as follows:
\begin{equation}\label{eq}
	\eta_{i} = \dfrac{d_{i2}}{8\pi\epsilon_{0}\chi_{2,2}^{2D}}\left(P(n_{ph}\ne 0)-P(n_{ph}=0) \right) 
\end{equation}
with $i=1,2$ standing for the $a$ and $b$ axis, $d_{12}$ and $d_{22}$ are the piezoelectric constants of the FE phase and $\chi_{2,2}^{2D}$ the 2D dielectric susceptibilities tensor component along the polar axis. We compute $\chi_{2}^{2D}$ and $d_{ij}$ for the FE phase of each NbOX\ensuremath{_{2}}, using DFPT as implemented in the Abinit code. Note that the $d_{ij}$ tensor is not a direct output of Abinit.  We rather construct it using the relation $e_{ij}= d_{ik}c_{kj}$, from the proper piezoelectric tensor $e_{ij}$ and the elastic constant $c_{ij}$ that are outputs directly from DFPT calculations. 
The obtained values are plotted on Fig. \ref{f3e}b, showing that the converse piezoelectric accounts for most of the photoinduced strains. Note that the model presented in Eq.\ref{eq} can also  be applied to the AFE phases given the fact that they are locally piezoelectric. To demonstrate that, we adapt Eq.\ref{eq} to the AFE phase by substituting the FE polarization with the local polarization of the AFE phase’s polar sublattice, while retaining the FE piezoelectric coefficient $d_{i2}$. The results presented in the SM \cite{xx} show good agreement with the DFT data for the AFE phase. This is not surprising because, although the local polarization in the AFE differs from that of the FE, the light-induced change in polarization is identical in both phases within the linear regime (See SM \cite{xx}). Also, as shown on Fig\ref{f3e}a, the light induced strain in both phases are quantitatively similar. These findings mean that, despite the lack of global piezoelectricity, the observed photostrictive behavior in the AFE phase can be attributed to a “local converse piezoelectric effect.”

\section{Conclusion}
In this work, we present first-principles investigation of the impact of optical excitation on the layered 2D NbOX\ensuremath{_{2}} materials. We demonstrate that illumination not only induces a FE to PE transition in these 2D systems but also an AFE to PE transition. Additionally, we present results on the photostrictive behavior in 2D NbOX\ensuremath{_{2}}, showing an enhanced behavior compared to bulk FE such as BiFeO\ensuremath{_{3}}, BaTiO\ensuremath{_{3}} and PbTiO\ensuremath{_{3}}. Notably, this significant photostrictive response can be realized at carrier densities that are experimentally accessible. Our study establishes 2D NbOX\ensuremath{_{2}} as promising candidates for applications in remote actuation, sensing, and non-volatile optoelectronic devices.

	\begin{acknowledgments}
This work is supported by the Grant No. MURI ETHOS W911NF-21-2-0162 from Army Research Office (ARO), the ARO Grant No. W911NF-21-1-0113, and the Vannevar Bush Faculty Fellowship (VBFF) Grant No. N00014-20-1-2834 from the Department of Defense. C.P. acknowledges support from the Air Force Office of  Scientific Research through Award No. FA9550-24-1-0263. This research is also supported by the Arkansas High Performance Computing Center, which is funded through multiple National Science Foundation grants and the Arkansas Economic Development Commission.
	\end{acknowledgments}
	
	\bibliography{Manuscript}
	
\end{document}